\documentclass{iau_FM}
\usepackage{graphicx}

\usepackage{soul} % For highlighting

\title[From NVSS to Future Polarisation Surveys] %% give here short title %%
{From the NVSS RM Catalogue to \\ Future Polarisation Surveys}

\author[Y.\ K.\ Ma \etal]   %% give here short author list %%
{Yik Ki Ma$^1$
  \thanks{Member of the International Max Planck Research School (IMPRS) for Astronomy and Astrophysics at the Universities of Bonn and Cologne},
S.\ A.\ Mao$^1$,
Jeroen Stil$^2$,
Aritra Basu$^{3,1}$, \\
Jennifer West$^4$,
Carl Heiles$^5$,
Alex S.\ Hill$^{6,7,8}$,
\and S.\ K.\ Betti$^9$}

\affiliation{
$^1$Max-Planck-Institut f\"{u}r Radioastronomie, Auf dem H\"{u}gel 69, 53121 Bonn, Germany \\ email: {\tt ykma@mpifr-bonn.mpg.de} \\[\affilskip]
$^2$Department of Physics and Astronomy, University of Calgary, 2500 University Drive NW, Calgary, AB T2N 1N4, Canada \\[\affilskip]
$^3$Fakult\"{a}t f\"{u}r Physik, Universit\"{a}t Bielefeld, Postfach 100131, 33501 Bielefeld, Germany \\[\affilskip]
$^4$Dunlap Institute for Astronomy and Astrophysics, The University of Toronto, 50 St.\ George Street, Toronto, ON M5S 3H4, Canada \\[\affilskip]
$^5$Department of Astronomy, University of California, Berkeley, CA 94720-3411, USA \\[\affilskip]
$^6$Department of Physics and Astronomy, The University of British Columbia, Vancouver, BC V6T 1Z1, Canada \\[\affilskip]
$^7$Space Science Institute, Boulder, CO, USA \\[\affilskip]
$^8$National Research Council Canada, Herzberg Program in Astronomy and Astrophysics, Dominion Radio Astrophysical Observatory, PO Box 248, Penticton, BC V2A 6J9, Canada \\[\affilskip]
$^9$Department of Astronomy, University of Massachusetts, 710 North Pleasant Street, Amherst, MA 01003-9305, USA
}

\pubyear{2018}
\jname{Astronomy in Focus, Volume ---} 
\editors{---, ed.}
\begin{document}

\maketitle

\begin{abstract}

With rotation measure (RM) towards 37,543 polarised sources, the \cite{taylor09} RM catalogue has been widely exploited in studies of the foreground magneto-ionic media. However, due to limitations imposed by observations in survey mode in the narrowband era, the listed RM values are inevitably affected by various systematic effects. With new Karl G.\ Jansky Very Large Array (VLA) broadband spectro-polarimetric observations at L-band, we set off to observationally examine the robustness of the Taylor catalogue. This would facilitate combinations and comparisons of it with results from current and future polarisation surveys such as Polarization Sky Survey of the Universe's Magnetism (POSSUM), VLA Sky Survey (VLASS), and the eventual Square Kilometre Array (SKA). Our on-axis pointed observations, in conjunction with simulations, allowed us to estimate the impact of off-axis polarisation leakage on the measured RM values. This demonstrates the importance to properly calibrate for the off-axis leakage terms in future all-sky polarisation surveys, in order to obtain high fidelity polarisation information from sources down to low fractional polarisation.

\end{abstract}

\firstsection % if your document starts with a section,
\section{Introduction}

Magnetic fields are known to be essential for the astrophysics in, for example, star formation, propagation of cosmic rays, and outflows and evolution of galaxies \cite[(see review by Beck \& Wielebinski 2013; Beck 2016)]{beck13, beck16}. One way to probe astrophysical magnetic fields is by quantifying the amount of Faraday rotation experienced by the linear-polarised synchrotron emission from background extragalactic sources (EGSs) in radio regime. This process changes the polarisation position angle (PA; [rad]) by
\begin{equation}
\Delta{\rm PA} = \left[ 0.81 \int_\ell^0 n_e(s) B_\parallel(s)\,{\rm d}s \right] \cdot \lambda^2 \equiv {\rm RM} \cdot \lambda^2{\rm ,}
\end{equation}
where $\ell$ [pc] is the (physical) distance to the background source, $n_e$ [${\rm cm}^{-3}$] is the thermal electron density, $B_\parallel$ [$\mu{\rm G}$] is the magnetic field strength along the line of sight ($s$ [pc]; increases away from the observer), $\lambda$ [m] is the wavelength of the polarised emission in question, and RM [${\rm rad\,m}^{-2}$] is the rotation measure of the background source. In other words, the average magnetic field strength (and direction) along the line of sight, weighted by $n_e$, is imprinted in the RM value of the polarised background source. This can be exploited in so-called RM-grid experiments, where RM measurements of numerous polarised background sources are performed to reveal the magnetic field strengths and structures in foreground astrophysical sources \cite[(e.g.\ Mao \etal\ 2008; Mao \etal\ 2010; Harvey-Smith \etal\ 2011; Van Eck \etal\ 2011; Gie\ss\"ubel \etal\ 2013; Kaczmarek \etal\ 2017; Mao \etal\ 2017)]{mao08,mao10,harveysmith11,vaneck11,giessuebel13,kaczmarek17,mao17}.

The largest RM catalogue to date is the \cite{taylor09} catalogue, in which the RM values were determined using narrowband Very Large Array (VLA) data at 1364.9 and 1435.1\,MHz (with bandwidths of 42\,MHz each) in 1990s from the NRAO VLA Sky Survey \cite[(NVSS; Condon \etal\ 1998)]{condon98}. With reported values through 37,543 lines of sight north of $\delta = -40^\circ$ (at a source density of about $1\,{\rm deg}^{-2}$), the Taylor catalogue allows studies of cosmic magnetism in astrophysical structures with angular sizes larger than several degrees \cite[(e.g.\ Harvey-Smith \etal\ 2011; Stil \etal\ 2011; Purcell \etal\ 2015)]{harveysmith11,stil11,purcell15}. While on-going broadband polarisation surveys such as Polarization Sky Survey of the Universe's Magnetism \cite[(POSSUM; Gaensler \etal\ 2010)]{gaensler10} and VLA Sky Survey \cite[(VLASS; Myers \etal\ 2014)]{myers14}, as well as the eventual Square Kilometre Array (SKA), will grant us much higher RM densities over the entire sky (see S.\ A.\ Mao in this volume), the Taylor catalogue will still remain a unique window to the magnetised Universe by complementing POSSUM and SKA in sky domain and VLASS in frequency domain. Comparisons between the Taylor catalogue and future survey results will also allow studies of polarisation time variabilities over more than 20 years.

Given the significance of the Taylor catalogue, we have investigated its robustness using new broadband data from the Karl G.\ Jansky VLA. The full results on $n\pi$-ambiguity and off-axis polarisation leakage in the Taylor catalogue, as well as Faraday complexities and potential RM time variabilities of our target sources, are reported in Ma \etal\ (submitted). In this proceedings, we focus on the implications of the off-axis instrumental polarisation leakage in the Taylor catalogue. The observation details are described in Section~2. The effects of $n\pi$-ambiguity and off-axis leakage are discussed in Sections~3 and 4. We draw the conclusion and remark on future polarisation surveys in Section~5.

\section{New Broadband Observations}

From the Taylor catalogue, we selected a sample of 23 sources for our new spectro-polarimetric observations. These sources have high RM magnitudes of $\gtrsim 300\,{\rm rad\,m}^{-2}$\footnote[2]{Except for NVSS J234033$+$133300, which has ${\rm RM}_{\rm Taylor} = +56.7 \pm 6.3\,{\rm rad\,m}^{-2}$.}, have Galactic latitude of $|b| > 10^\circ$, and have NVSS flux densities higher than $100\,{\rm mJy}$. The unusually high $|{\rm RM}|$ values of these sources make them prime $n\pi$-ambiguity candidates in the Taylor catalogue (see Ma \etal\ submitted).

New broadband data were obtained using Karl G.\ Jansky VLA in L-band (1--2\,GHz) in D-array configuration on 2014 July 03. The total integration time per source is about 3--4 minutes, with our targets placed on the pointing axis of the antennae to ensure no off-axis instrumental effects are present in our data. Standard flagging and calibration procedures were applied to the data, including one iteration of phase self calibration on the target sources. From this, we formed two sets of Stokes \textit{I}, \textit{Q}, and \textit{U} images --- ``full band images'' utilising the entire usable band, binning 4\,MHz of visibility data for each step in frequency, and ``NVSS band images'' formed with nearly identical frequency coverages as that of the NVSS survey data used by the Taylor catalogue, albeit some flaggings were done to our data in those frequency ranges. On one hand, we use the full band images, combined with the RM-Synthesis algorithm \cite[(Brentjens \& de Bruyn 2005)]{brentjens05}, to obtain the Faraday depth (FD; the generalisation of RM, see e.g.\ \cite[Brentjens \& de Bruyn 2005]{brentjens05}, Ma \etal\ submitted) of our sources free of $n\pi$-ambiguity. On the other hand we use the NVSS band images to make direct comparisons with the Taylor catalogue results.

\section{$n\pi$-ambiguity in the Taylor Catalogue \label{sec:npi}}

We compared the FD from RM-Synthesis on our full-band images with the RM from the Taylor catalogue (${\rm RM}_{\rm Taylor}$), and found that the difference between these two sets of values for nine of our targets are consistent with that expected from $n\pi$-ambiguity in the Taylor catalogue. By comparing the statistical properties of these sources versus that of sources with reliable RM values, we estimate that there may be more than 50 $n\pi$-ambiguity sources out of the 37,543 Taylor catalogue sources.

  \begin{figure}[h]
\begin{center}
  \includegraphics[width=2.63in]{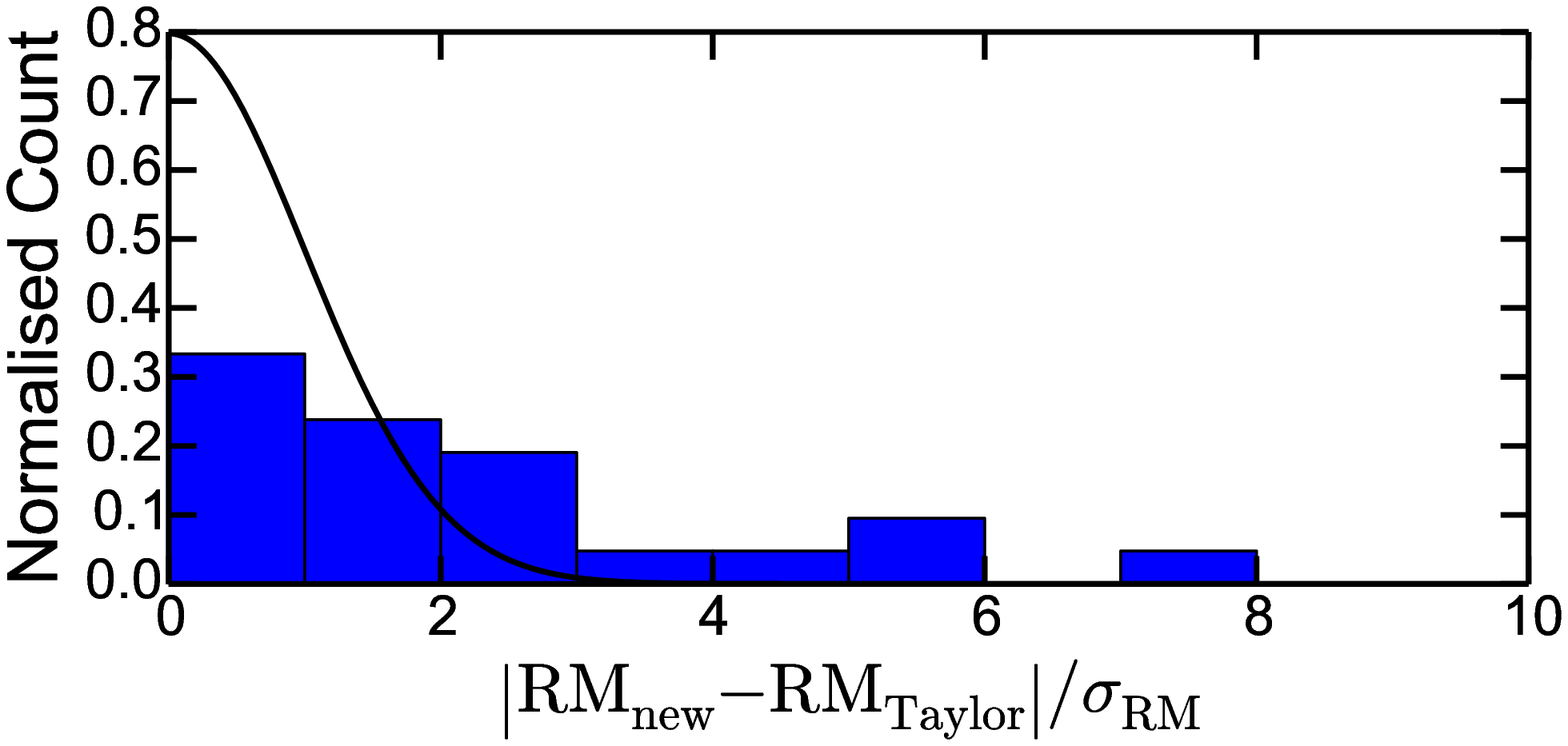} 
 \includegraphics[width=2.63in]{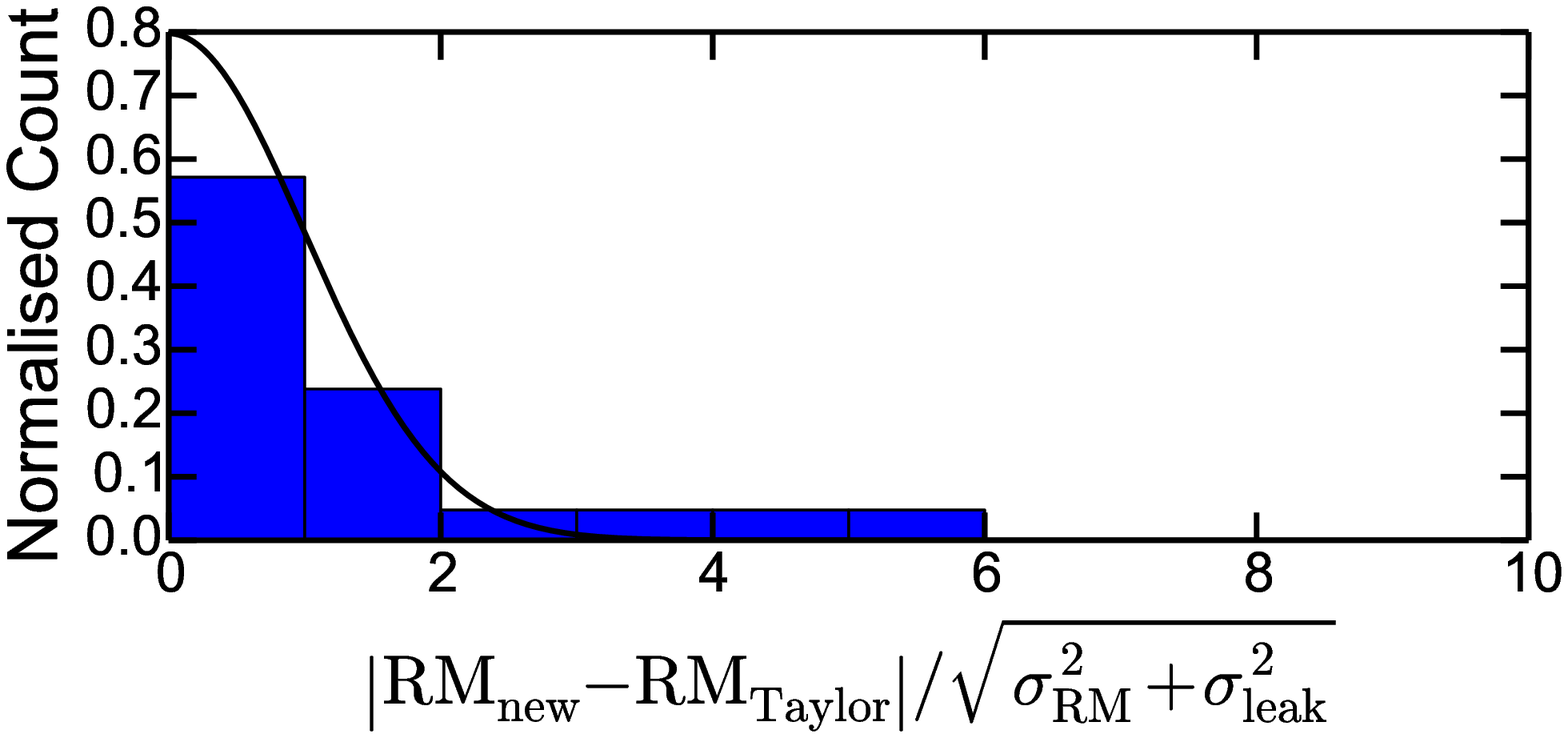} 
 \caption{Comparisons between our ${\rm RM}_{\rm new}$ and ${\rm RM}_{\rm Taylor}$. \textbf{(Left)} Histogram of RM differences in units of RM uncertainties of the observations($\sigma_{\rm RM}$); \textbf{(Right)} The same histogram as the left panel, except that we have accounted for the extra RM uncertainty due to off-axis leakage ($\sigma_{\rm leak}$). The black line in both panels shows a folded normal distribution, which is the expected profile of the histogram if the RM differences are purely due to the included uncertainties. \label{fig:rmdiff}}
\end{center}
\end{figure}

\section{Polarisation Discrepancies due to Off-axis Polarisation Leakage}

Through careful comparisons between our results and those from the Taylor catalogue, we noted discrepancies in polarisation properties from the two studies. Firstly, from RM-Synthesis analysis in Section~\ref{sec:npi}, we found that two of our target sources are unpolarised ($\lesssim 0.07\,\%$ at $6\sigma$), though they are listed as $\approx 0.5\,\%$ polarised in the Taylor catalogue with signal-to-noise of over 30 in polarisation. Secondly, the RM derived from our new NVSS band images (${\rm RM}_{\rm new}$) and ${\rm RM}_{\rm Taylor}$ do not match within measurement uncertainties (Figure~\ref{fig:rmdiff} Left), after correcting both for $n\pi$-ambiguity using the broadband FD obtained in Section~\ref{sec:npi}.

These discrepant polarisation properties are mainly due to off-axis polarisation leakage in the Taylor catalogue. A simulation was conducted to quantify its effect on RM measurements in the Taylor catalogue. Using the Taylor catalogue listed values (specifically, total intensity, polarised intensity, and ${\rm RM}_{\rm Taylor}$) as inputs, we injected an artificial polarisation leakage signal with amplitudes fixed at $0.5\,\%$ of the total intensity to the source polarisation signal, and compared the derived RM with and without this leakage effect (see Ma \etal\ submitted for details of the simulation). We found that on average, the effect of the uncorrected off-axis polarisation leakage can be taken into account by increasing the listed RM uncertainties in the Taylor catalogue by $10\,\%$. By incorporating this extra RM uncertainty to our sources, we find that the discrepancies between ${\rm RM}_{\rm new}$ and ${\rm RM}_{\rm Taylor}$ can be mostly explained (Figure~\ref{fig:rmdiff} Right), except for three sources where significant RM differences still remain. This could be due to true RM time variabilities which will be investigated in a future study (Ma \etal\ in prep).
 
\section{Conclusion and Outlook to Future Polarisation Surveys}
 
In this proceedings, we reported our findings on the robustness of RM values reported in the Taylor catalogue. Our results suggest that the $n\pi$-ambiguity may affect the RM values of more than 50 out of the total of 37,543 sources in the Taylor catalogue. Furthermore, the uncorrected off-axis polarisation leakage in the NVSS data results in an extra $10\,\%$ in Taylor RM uncertainties. These effects must be taken into account in future studies utilising this RM catalogue.
 
We demonstrated the effect of uncorrected off-axis instrumental polarisation on RM measurements in the specific case of the Taylor catalogue. This same instrumental effect can also impact future broadband polarisation surveys, manifested as an artificial signal in Faraday spectrum near $0\,{\rm rad\,m}^{-2}$ (see J.\ Stil in this volume). To ensure high fidelity in Faraday spectra, particularly from sources with low fractional polarisation, current and future polarisation surveys must properly characterise and remove the off-axis polarisation leakage terms.

\end{document}